\documentstyle[12pt]{article}

\input{tcilatex}

\begin{document}

\title{A pseudo - perturbative expansions method;  nonpolynomial, cutoff - Coulomb,
and Coulomb plus logarithmic potentials}
\author{Omar Mustafa and Maen Odeh \\
Department of Physics, Eastern Mediterranean University\\
G. Magusa, North Cyprus, Mersin 10 - Turkey\\
email: omar.mustafa@emu.edu.tr\\
\ \ \ \ \ \ \ \ \ \ maen.odeh@emu.edu.tr\newline
}
\maketitle

\begin{abstract}
{\small We propose a new analytical method to solve for nonexactly soluble
Schr\"{o}dinger equation via expansions through some existing quantum
numbers. Successfully, it is applied to the rational nonpolynomial
oscillator potential. Moreover, a conclusion reached by Scherrer et al. [2],
via matrix continued fractions method, that the shifted large N expansion
method leads to dubious accuracies is investigated. The cutoff - Coulomb and
Coulomb plus logarithmic potentials are also investigated.}
\end{abstract}

\newpage

\newpage

\section{Introduction}

The importance of rational nonpolynomial oscillator (NPO) potential,\newline
\begin{equation}
V(q)=a_o q^2+ \frac{aq^2}{1+bq^2}~,~~b>0,
\end{equation}
\newline
arises in nonlinear Lagrangian field theory, laser theory, and elementary
particle physics [1-4]. Only a class of exact analytical solutions for a
certain parameter dependence $a=a(b)$ obtains [3,5-7]. Hence, it has been a
subject of several investigations [1-15] ( exhaustive lists of these could
be found in Ref.s[2,15]).

On the other hand, the cutoff - Coulomb potential\newline
\begin{equation}
V(q)=-\frac{1}{q+c}~~,~~c>0,
\end{equation}
\newline
with the truncation parameter $c$, avoids the singularity at $q=0$ in the
Coulomb potential ( the crux of divergence difficulties in quantum field
theory [16-18]). It has been suggested [18] that if gravitational
interactions of elementary particles are taken into acount, there would be a
gravitational cutoff of Coulomb interaction. Equation (2) represents a
nonrelativistic expression of this idea. It also serves as an approximation
to the potential due smeared charge rather than a point charge.

The Schr\"odinger equation with such interactions, (1) and (2), belongs to
the class of quantum mechanical systems which are nonexactly soluble in
general. However, the solutions of exactly solvable potentials ( an
interesting field of mathematical physics in itself [19,20]) can be used in
perturbation and pseudoperturbation theories, or they can be combined with
numerical calculations. Nevertheless, in the simplest case, analytical
calculations can aid numerical studies in areas where numerical techniques
might not be safely controlled. For example, when bound - state wave
functions with arbitrary nodal zeros are required for certain singular
potentials (a next level of complexity), analytical solutions can supply a
basis for numerical calculations. Moreover, in many problems the Hamiltonian
does not contain any physical parameter suitable for a perturbation
expansion treatment. More often, the Hamiltonian contains physical
parameters, but, typically, zeroth - order solutions for special values of
these are not tractable or good starting approximations. Alternatively, one
would resort to apparently artificial conversions to perturbation problems
which have been shown to make progress [21-35].

The three - dimensional (3D) spherically symmetric NPO potential has been
investigated [8] by means of the shifted large - N expansion technique
(SLNT). However, Scherrer and co-workers [2] have employed a matrix
continued fractions (MCF) method and concluded that SLNT [8] leads to
dubiously accurate results in the critical range $a^{1/2}/b \approx 0.1-30$.
Handy et al. [14] have used the eigenvalue moment method (EMM) to obtain
upper and lower energy bounds, for $l \neq 0$, and compared their results
with those of Roy et al. [12] by the suppersymmetric quantum mechanics
(SSQM). Singh et al [17] have used a numerical integration method (NIM) to
find the eigenvalues for the 3D cutoff - Coulomb potential. Hall and Saad
[35] have used a smooth transformation (STM) and a numerical integration
methods to obtain bound - states for the Coulomb plus a logarithmic
perturbation term. Hence, apart from those of SLNT, results from
convincingly powerful methods exist for comparison purposes.

Recently, we have introduced a pseudoperturbative shifted - $l$ ( $l$ is the
angular momentum quantum number) expansion technique ( PSLET) to solve for
nodeless states of Schr\"odinger equation. It simply consists of using $1/%
\bar{l}$ as a pseudoperturbation parameter, where $\bar{l}=l-\beta$ and $%
\beta$ is a suitable shift. The shift $\beta$ is vital for it removes the
poles that would emerge, at lowest orbital states with $l$=0, in our
proposed expansions below. Our analytical, or often semianalytical,
methodical proposal PSLET has been successfully applied to quasi -
relativistic harmonic oscillator [34], spiked harmonic oscillator [32], and
to anharmonic oscillators [35] potentials.

Encouraged by its satisfactory performance in handling nodeless states, we
generalize PSLET recipe ( in section 2) for states with arbitrary number of
nodal zeros, $k \geq 0$. Moreover, in the underlying "radical" time -
independent radial Schr\"odinger equation, in $\hbar=m=1$ units,\newline
\begin{equation}
\left[-\frac{1}{2}\frac{d^{2}}{dq^{2}}+\frac{l(l+1)}{2q^{2}}+V(q)\right]
\Psi_{k,l}(q)=E_{k,l}\Psi_{k,l}(q),
\end{equation}
\newline
the isomorphism between orbital angular momentum $l$ and dimensionality $D$
invites interdimensional degeneracies to obtain [36-39] (for more details
the reader may refer to ref.s [36,38]). Hence, the symmetry of an attendant
problem obviously manifests the admissibility of the quantum number $l$: In
one-dimension (1D), $l$ specifies parity, $(-1)^{l+1}$, with the permissible
values -1 and/or 0 ( even and/or odd parity, respectively) where $%
q=x\in(-\infty,\infty)$. For two-dimensional (2D) cylindrically symmetric
Schr\"odinger equation one sets $l=|m|-1/2$, where m is the magnetic quantum
number and $q=(x^2+y^2)^{1/2}\in(0,\infty)$. Finally, for three-dimensional
(3D) spherically symmetric Schr\"odinger equation, $l$ denotes the angular
momentum quantum number with $q=(x^2+y^2+z^2)^{1/2}\in(0,\infty)$.

In section 3, we investigate PSLET recipe and consider, for the sake of
diversity, the potentials; (i) $V(q)=A^2q^2/2$, the harmonic osillator ( the
limit of (1) when $b \longrightarrow \infty$ and/or $a \longrightarrow 0$,
with $a_o=A^2/2$), (ii) $V(q)=-1/q$, the Coulomb ( the limit of (2) when $c
\longrightarrow 0$, (iii) the NPO (2), (iv) the cutoff - Coulomb (2), and
(v) the Coulomb perturbed by a logarithmic term, which has no experimental
evidence, to the best of our knowledge, thus our calculations are only of
academic interest. The last section is reserved for summary and remarks.

\section{The generalization of PSLET}

The construction of our PSLET starts with shifting the angular momentum in
(3) to obtain\newline
\begin{equation}
\left\{-\frac{1}{2}\frac{d^{2}}{dq^{2}}+\frac{\bar{l}^{2}+(2\beta+1)\bar{l}
+\beta(\beta+1)}{2q^{2}}+\frac{\bar{l}^2}{Q}V(q) \right\} \Psi_{k,l}
(q)=E_{k,l}\Psi_{k,l}(q),
\end{equation}
\newline
where Q is a constant that scales the potential $V(q)$ at large - $l$ limit
( the pseudoclassical limit [36]) and is set, for any specific choice of $l$
and $k$, equal to $\bar{l}^2$ at the end of the calculations. Next, we shift
the origin of the coordinate system through $x=\bar{l}^{1/2}(q-q_{o})/q_{o}$%
, where $q_o$ is currently an arbitrary point to be determined below.
Expansions about this point, $x=0$ (i.e. $q=q_o$), yield\newline
\begin{equation}
\frac{1}{q^{2}}=\sum^{\infty}_{n=0} (-1)^{n}~ \frac{(n+1)}{q_{o}^{2}} ~
x^{n}~\bar{l}^{-n/2},
\end{equation}
\newline
\begin{equation}
V(x(q))=\sum^{\infty}_{n=0}\left(\frac{d^{n}V(q_{o})}{dq_{o}^{n}}\right) 
\frac{(q_{o}x)^{n}}{n!}~\bar{l}^{-n/2}.
\end{equation}
\newline
Obviously, the expansions in (5) and (6) localize the problem at an
arbitrary point $q_o$ and the derivatives, in effect, contain information
not only at $q_o$ but also at any point on $q$-axis, in accordance with
Taylor's theorem. It is then convenient to expand $E_{k,l}$ as\newline
\begin{equation}
E_{k,l}=\sum^{\infty}_{n=-2}E_{k,l}^{(n)}~\bar{l}^{-n}.
\end{equation}
\newline
Equation (4) thus becomes\newline
\begin{equation}
\left[-\frac{1}{2}\frac{d^{2}}{dx^{2}}+\frac{q_{o}^{2}}{\bar{l}} \tilde{V}%
(x(q))\right] \Psi_{k,l}(x)=\frac{q_{o}^2}{\bar{l}}E_{k,l}\Psi_{k,l}(x),
\end{equation}
\newline
with\newline
\begin{eqnarray}
\frac{q_o^2}{\bar{l}}\tilde{V}(x(q))&=&q_o^2\bar{l} \left[\frac{1}{2q_o^2}+%
\frac{V(q_o)}{Q}\right] +\bar{l}^{1/2}B_1 x+ B_2 x^2+\frac{(2\beta+1)}{2} 
\nonumber \\
&+&(2\beta+1)\sum^{\infty}_{n=1}(-1)^n \frac{(n+1)}{2}x^n \bar{l}^{-n/2}
+\sum^{\infty}_{n=3}B_n x^n\bar{l}^{-(n-2)/2}  \nonumber \\
&+& \beta(\beta+1)\sum^{\infty}_{n=0}(-1)^n\frac{(n+1)}{2}x^n\bar{l}%
^{-(n+2)/2},
\end{eqnarray}
\newline
\begin{equation}
B_n=(-1)^n \frac{(n+1)}{2} +\left(\frac{d^n V(q_o)}{dq_o^n}\right)\frac{%
q_o^{n+2}}{n! Q}.
\end{equation}
\newline
Equation (8), along with (9) and (10), is evidently the one - dimensional
Schr\"odinger equation for a perturbed harmonic oscillator\newline
\begin{equation}
\left[-\frac{1}{2}\frac{d^2}{dx^2}+\frac{1}{2}w^2 x^2 +\varepsilon_o +P(x)%
\right]X_{k}(x)=\lambda_{k}X_{k}(x),
\end{equation}
\newline
where $w^2=2B_2$, 
\begin{equation}
\varepsilon_o =\bar{l}\left[\frac{1}{2}+\frac{q_o^2 V(q_o)}{Q}\right] +\frac{%
2\beta+1}{2}+\frac{\beta(\beta+1)}{2\bar{l}},
\end{equation}
\newline
and $P(x)$ represents the remaining terms in eq.(9) as infinite power series
perturbations to the harmonic oscillator. One would then imply that\newline
\begin{eqnarray}
\lambda_{k}&=&\bar{l}\left[\frac{1}{2}+\frac{q_o^2 V(q_o)}{Q}\right] +\left[%
\frac{2\beta+1}{2}+(k+\frac{1}{2})w\right]  \nonumber \\
&+&\frac{1}{\bar{l}}\left[\frac{\beta(\beta+1)}{2}+\lambda_{k}^{(0)}\right]
+\sum^{\infty}_{n=2}\lambda_{k}^{(n-1)}\bar{l}^{-n},
\end{eqnarray}
\newline
and 
\begin{equation}
\lambda_{k} = q_o^2 \sum^{\infty}_{n=-2} E_{k,l}^{(n)} \bar{l}^{-(n+1)}.
\end{equation}
\newline
Hence, equations (13) and (14) yield\newline
\begin{equation}
E_{k,l}^{(-2)}=\frac{1}{2q_o^2}+\frac{V(q_o)}{Q}
\end{equation}
\newline
\begin{equation}
E_{k,l}^{(-1)}=\frac{1}{q_o^2}\left[\frac{2\beta+1}{2} +(k +\frac{1}{2})w%
\right]
\end{equation}
\newline
\begin{equation}
E_{k,l}^{(0)}=\frac{1}{q_o^2}\left[ \frac{\beta(\beta+1)}{2}
+\lambda_{k}^{(0)}\right]
\end{equation}
\newline
\begin{equation}
E_{k,l}^{(n)}=\lambda_{k}^{(n)}/q_o^2 ~~;~~~~n \geq 1.
\end{equation}
\newline
Where $q_o$ is chosen to minimize $E_{k,l}^{(-2)}$, i. e.\newline
\begin{equation}
\frac{dE_{k,l}^{(-2)}}{dq_o}=0~~~~ and~~~~\frac{d^2 E_{k,l}^{(-2)}}{dq_o^2}%
>0.
\end{equation}
\newline
Hereby, $V(q)$ is assumed to be well behaved so that $E_{k,l}^{(-2)}$ has a
minimum $q_o$ and there are well - defined bound - states. Equation (19) in
turn gives, with $\bar{l}=\sqrt{Q}$, 
\begin{equation}
l-\beta=\sqrt{q_{o}^{3}V^{^{\prime}}(q_{o})}.
\end{equation}
\newline
Consequently, the second term in Eq.(9) vanishes and the first term adds a
constant to the energy eigenvalues. It should be noted that the energy term $%
\bar{l}^2E_{k,l}^{(-2)}$ corresponds roughly to the energy of a classical
particle with angular momentum $L_z$=$\bar{l}$ executing circular motion of
radius $q_o$ in the potential $V(q_o)$. It thus identifies the zeroth -
order approximation, to all eigenvalues, as a classical approximation and
the higher - order corrections as quantum fluctuations around the minimum $%
q_o$, organized in inverse powers of $\bar{l}$. The next correction to the
energy series, $\bar{l}E_{k,l}^{(-1)}$, consists of a constant term and the
exact eigenvalues of the harmonic oscillator $w^2x^2/2$.The shifting
parameter $\beta$ is determined by choosing $\bar{l}E_{k,l}^{(-1)}$=0. This
choice is physically motivated. In addition to its vital role in removing
the singularity at $l=0$, it also requires the agreements between PSLET
eigenvalues and eigenfunctions with the exact well known ones for the
harmonic oscillator and Coulomb potentials. Hence\newline
\begin{equation}
\beta=-\left[\frac{1}{2}+(k+\frac{1}{2})w\right]~,~~ w=\sqrt{3+\frac{q_o
V^{^{\prime\prime}}(q_o)}{V^{^{\prime}}(q_o)}}
\end{equation}
\newline
where primes of $V(q_o)$ denote derivatives with respect to $q_o$. Then
equation (9) reduces to\newline
\begin{equation}
\frac{q_o^2}{\bar{l}}\tilde{V}(x(q))= q_o^2\bar{l}\left[\frac{1}{2q_o^2}+%
\frac{V(q_o)}{Q}\right]+ \sum^{\infty}_{n=0} v^{(n)}(x) \bar{l}^{-n/2},
\end{equation}
\newline
where\newline
\begin{equation}
v^{(0)}(x)=B_2 x^2 + \frac{2\beta+1}{2},
\end{equation}
\newline
\begin{equation}
v^{(1)}(x)=-(2\beta+1) x + B_3 x^3,
\end{equation}
\newline
\begin{eqnarray}
v^{(n)}(x)&=&B_{n+2}~ x^{n+2}+(-1)^n~ (2\beta+1)~ \frac{(n+1)}{2}~ x^n 
\nonumber \\
&+&(-1)^{n}~ \frac{\beta(\beta+1)}{2}~ (n-1)~ x^{(n-2)}~,~~ n \geq 2.
\end{eqnarray}
\newline
Equation (8) thus becomes\newline
\begin{equation}
\left[-\frac{1}{2}\frac{d^2}{dx^2} + \sum^{\infty}_{n=0} v^{(n)} \bar{l}%
^{-n/2}\right]\Psi_{k,l} (x)= \left[\sum^{\infty}_{n=1} q_o^2
E_{k,l}^{(n-1)} \bar{l}^{-n} \right] \Psi_{k,l}(x).
\end{equation}
\newline

Up to this point, one would conclude that the above procedure is nothing but
an imitation of the eminent shifted large-N expansion (SLNT) [24-29,35].
However, because of the limited capability of SLNT in handling large-order
corrections via the standard Rayleigh-Schr\"odinger perturbation theory,
only low-order corrections have been reported, sacrificing in effect its
preciseness. Therefore, one should seek for an alternative and proceed by
setting the wave functions with any number of nodes as \newline
\begin{equation}
\Psi_{k,l}(x(q)) = F_{k,l}(x)~ exp(U_{k,l}(x)).
\end{equation}
\newline
In turn, equation (26) readily transforms into the following Riccati
equation:\newline
\begin{eqnarray}
&&F_{k,l}(x)\left[-\frac{1}{2}\left(
U_{k,l}^{^{\prime\prime}}(x)+U_{k,l}^{^{\prime}}(x)
U_{k,l}^{^{\prime}}(x)\right) +\sum^{\infty}_{n=0} v^{(n)}(x) \bar{l}^{-n/2}
\right.  \nonumber \\
&&\left. -\sum^{\infty}_{n=1} q_o^2 E_{k,l}^{(n-1)} \bar{l}^{-n} \right]
-F_{k,l}^{^{\prime}}(x)U_{k,l}^{^{\prime}}(x)-\frac{1}{2}F_{k,l}^{^{\prime%
\prime}}(x)=0,
\end{eqnarray}
\newline
where the primes denote derivatives with respect to $x$. It is evident that
this equation admits solution of the form \newline
\begin{equation}
U_{k,l}^{^{\prime}}(x)=\sum^{\infty}_{n=0} U_{k}^{(n)}(x)~~\bar{l}^{-n/2}
+\sum^{\infty}_{n=0} G_{k}^{(n)}(x)~~\bar{l}^{-(n+1)/2},
\end{equation}
\newline
\begin{equation}
F_{k,l}(x)=x^k +\sum^{\infty}_{n=0}\sum^{k-1}_{p=0} a_{p,k}^{(n)}~~x^p~~\bar{%
l}^{-n/2},
\end{equation}
\newline
where\newline
\begin{equation}
U_{k}^{(n)}(x)=\sum^{n+1}_{m=0} D_{m,n,k}~~x^{2m-1} ~~~~;~~~D_{0,n,k}=0,
\end{equation}
\newline
\begin{equation}
G_{k}^{(n)}(x)=\sum^{n+1}_{m=0} C_{m,n,k}~~x^{2m}.
\end{equation}
\newline
Substituting equations (29) - (32) into equation (28) implies\newline
\begin{eqnarray}
&&F_{k,l}(x)\left[-\frac{1}{2}\sum^{\infty}_{n=0}\left(U_{k}^{(n)^{^{%
\prime}}} \bar{l}^{-n/2} + G_{k}^{(n)^{^{\prime}}} \bar{l}^{-(n+1)/2}\right)
\right.  \nonumber \\
&-&\left.\frac{1}{2} \sum^{\infty}_{n=0} \sum^{n}_{m=0} \left(
U_{k}^{(m)}U_{k}^{(n-m)} \bar{l}^{-n/2} +G_{k}^{(m)}G_{k}^{(n-m)} \bar{l}%
^{-(n+2)/2} \right. \right.  \nonumber \\
&+&\left.\left.2 U_{k}^{(m)}G_{k}^{(n-m)} \bar{l}^{-(n+1)/2}\right)
+\sum^{\infty}_{n=0}v^{(n)} \bar{l}^{-n/2} -\sum^{\infty}_{n=1} q_o^2
E_{k,l}^{(n-1)} \bar{l}^{-n}\right]  \nonumber \\
&-&F_{k,l}^{^{\prime}}(x)\left[\sum^{\infty}_{n=0}\left(U_{k}^{(n)}\bar{l}%
^{-n/2} + G_{k}^{(n)} \bar{l}^{-(n+1)/2}\right)\right]-\frac{1}{2}%
F_{k,l}^{^{\prime\prime}}(x) =0
\end{eqnarray}
\newline
The above procedure obviously reduces to the one described by Mustafa and
Odeh [33,34,36,37], for $k=0$. Moreover, the solution of equation (33)
follows from the uniqueness of power series representation. Therefore, for a
given $k$ we equate the coefficients of the same powers of $\bar{l}$ and $x$%
, respectively. For example, when $k=1$ one obtains\newline
\begin{equation}
D_{1,0,1}=-w,~~~ U_{1}^{(0)}(x) =-~w~x,
\end{equation}
\newline
\begin{equation}
C_{1,0,1}=-\frac{B_{3}}{w},~~~~a_{0,1}^{(1)}=-\frac{C_{0,0,1}}{w},
\end{equation}
\newline
\begin{equation}
C_{0,0,1}=\frac{1}{w}\left(2C_{1,0,1}+2\beta+1\right),
\end{equation}
\newline
\begin{equation}
D_{2,2,1}=\frac{1}{w}\left(\frac{C_{1,0,1}^2}{2}-B_{4}\right),
\end{equation}
\newline
\begin{equation}
D_{1,2,1}=\frac{1}{w}\left(\frac{5}{2}~D_{2,2,1}+C_{0,0,1}~C_{1,0,1} -\frac{3%
}{2}(2\beta+1)\right),
\end{equation}
\newline
\begin{equation}
E_{1,l}^{(0)} = \frac{1}{q_o^2}\left(\frac{\beta(\beta+1)}{2}+
a_{0,1}^{(1)}~C_{1,0,1}-\frac{3~D_{1,2,1}}{2}-\frac{C_{0,0,1}^2}{2}\right),
\end{equation}
\newline
etc. Here, we reported the nonzero coefficients only. One can then calculate
the energy eigenvalues and eigenfunctions from the knowledge of $C_{m,n,k}$, 
$D_{m,n,k}$, and $a_{p,k}^{(n)}$ in a hierarchical manner. Nevertheless, the
procedure just described is suitable for a software package such as MAPLE to
determine the energy eigenvalue and eigenfunction corrections up to any
order of the pseudoperturbation series, (7) and (29)-(30).

Although the energy series, equation (7), could appear divergent, or, at
best, asymptotic for small $\bar{l}$, one can still calculate the
eigenenergies to a very good accuracy by forming the sophisticated [N,M]
Pad\'e approximation [21]\newline

\begin{center}
$P_{N}^{M}(1/\bar{l})=(P_0+P_1/\bar{l}+\cdots+P_M/\bar{l}^M)/ (1+q_1/\bar{l}%
+\cdots+q_N/\bar{l}^N)$
\end{center}

to the energy series (7). The energy series is calculated up to $%
E_{k,l}^{(8)}/\bar{l}^8$ by 
\begin{equation}
E_{k,l}=\bar{l}^{2}E_{k,l}^{(-2)}+E_{k,l}^{(0)}+\cdots +E_{k,l}^{(8)}/\bar{l}%
^8+O(1/\bar{l}^{9}),
\end{equation}
\newline
and with the $P_{4}^{4}(1/\bar{l})$ Pad\'e approximant it becomes\newline
\begin{equation}
E_{k,l}[4,4]=\bar{l}^{2}E_{k,l}^{(-2)}+P_{4}^{4}(1/\bar{l}).
\end{equation}
\newline
Our recipe is therefore well prescribed.

\section{Some applications}

We begin with the harmonic oscillator potential $V(q)=A^{2}q^{2}/2$ and find 
$q_o$ from (20), along with (21). Once $q_o$ is determined, Eq.(15) reads%
\newline
\begin{equation}
E_{k,l}^{(-2)}\bar{l}^2 = A \bar{l} ~~;~~\bar{l} =2k+l+3/2,
\end{equation}
\newline
\begin{equation}
E_{k,l}^{(-1)}\bar{l} = E_{k,l}^{(0)} = E_{k,l}^{(1)}\bar{l}^{-1} =
E_{k,l}^{(2)}\bar{l}^{-2} = \cdots = 0,
\end{equation}
\newline
Hence, the corresponding energies are \newline
\begin{equation}
E_{0,l} = A\left(l+\frac{3}{2}\right),
\end{equation}
\newline
the well known exact results. For $k=0$\newline
\begin{eqnarray}
U_{0,l}(x)&=&-\frac{1}{2}\left(y-\frac{y^2}{2}+\frac{y^3}{3}-\frac{y^4}{4} +%
\frac{y^5}{5}-\frac{y^6}{6}+\cdots\right)  \nonumber \\
&&+\bar{l}\left(y-\frac{y^2}{2}+\frac{y^3}{3}-\frac{y^4}{4}+\frac{y^5}{5} -%
\frac{y^6}{6}+\cdots\right)  \nonumber \\
&&-\bar{l}\frac{y^2}{2}-\bar{l}y~~;~~y=x\bar{l}^{-1/2}.
\end{eqnarray}
\newline
Obviously the terms between brackets in equation (44) are the infinite
geometric series expansions for $ln(1+y)$. Equation (44) can be recast as 
\newline
\begin{equation}
U_{0,l}(x) = ln(1+y)^{-1/2} + ln(1+y)^{\bar{l}} - \bar{l}y - \bar{l}\frac{y^2%
}{2},
\end{equation}
\newline
which in turn implies the exact wave functions\newline
\begin{equation}
\Psi_{0,l} (q) = N_{0,l} q^{\bar{l}-1/2} e^{-Aq^2 /2}~~~;~~~\bar{l}=Aq_o^2,
\end{equation}
\newline
where $N_{0,l}$ are the normalization constants.

Next we consider the Coulomb potential $V(q)=-1/q$. In this case, Eq.(15)
reads\newline
\begin{equation}
E_{k,l}^{(-2)}\bar{l}^2 = -\frac{1}{2\bar{l}^{2}}~~;~~\bar{l}=k+l+1=q_o^2.
\end{equation}
\newline
The reminder corrections are identically zeros, Eq.(43). Hence, one obtains
the eigenvalues\newline
\begin{equation}
E_{k,l} = \frac{-1}{2(k+l+1)^2},
\end{equation}
\newline
the well known exact energies for the Coulomb potential. For $k=0$\newline
\begin{equation}
U_{0,l}(x) = - \bar{l} y + ln(1+y)^{\bar{l}},
\end{equation}
\newline
which in turn implies the exact wave functions\newline
\begin{equation}
\Psi_{0,l} (q) = N_{0,l} q^{\bar{l}} e^{-\bar{l} q/q_o}.
\end{equation}
\newline

Now let us consider the NPO potential (1), for which Eq.(21) implies\newline
\begin{equation}
\beta=-\frac{1}{2}[1+(2k+1)w]~~;~~~~ w=2\sqrt{\frac{%
(1+3bq_o^2+3b^2q_o^4+b^3q_o^6+a)} {(1+2bq_o^2+b^2q_o^4+a)(1+bq_o^2)}}.
\end{equation}
\newline
In turn Eq.(20) reads\newline
\begin{equation}
l+\frac{1}{2}[1+(2k+1)w] =\frac{q_o^2}{1+bq_o^2} \sqrt{1+a+bq_o^2+b^2q_o^4}.
\end{equation}
\newline
Equation (43) is explicit in $q_o$ and evidently a closed form solution for $%
q_o$ is hard to find, though almost impossible. However, numerical solutions
are feasible. Once $q_o$ is determined the coefficients $C_{m,n}$ and $%
D_{m,n}$ are obtained in a sequel manner. Consequently, the eigenvalues,
Eq.(40), and eigenfunctions, Eqs.(29)-(32), are calculated in the same batch
for each value of $a$, $b$, and $l$.

In order to make remediable analysis of our results we have calculated the
first ten terms of the energy series. The effect of each term has been taken
into account. We have also computed the Pad\'{e} approximants $E[N,M]$ for $%
N=2,3,4$ and $M=2,3,4$. Therefore, the stability of the sequence of the
Pad\'{e} approximants was in point.

In tables 1 and 2 we list PSLET results $E_P$, Eq.(40), along with the [4,4]
Pad\'{e} approximants, Eqs.(41). The results of Roy et al.[12], via SSQM,
and Handy et al.[14], via EMM, are also displayed for comparison. In tables
3-5 we compare our results with those of Scherrer et al.[2], via MCF, and
Varshni [8], via SLNT.

In tables 1 and 2 we have observed that, if the last two digits are
neglected, the energy series Eq.(40) stabilizes at $E_5$ up to $E_{10}$.
Where $E_5$ and $E_{10}$ denote that the energy is computed by the first
five and first ten terms of the energy series, respectively. Yet for $l= 5 -
20$ it stabilizes at $E_3$ up to $E_{10}$. While the Pad\'{e} approximants
for $l= 5 - 20$ had no effect on the energy series their effect is not
dramatic for $l= 1 - 3$. Also it should be reported that the sequence of
Pad\'{e} approximants stabilizes at $E[3,3]$ for $l=1-3$ and at $E[2,2]$ for 
$l=5-20$. Therefore, one could confidently conclude that the results from
the Pad\'{e} approximants are exact provided that some uncertainty lies in
the last two digits for $l=1-3$. Nevertheless, our results do not contradict
with the upper and lower bounds computed from EMM [14]. To a satisfactory
extent they also agree with those from SSQM [12].

In tables 3 - 5 we list our results along with those of Scherrer [2], from
MCF, and Varshni [8], from SLNT. Our results compare better with those of
Scherrer than the results of SLNT. However, collecting only the first four
terms of the energy series Eq.(40) we found that PSLET results are in exact
accord with SLNT. Moreover, the trends of convergence of the energy series
and the sequence of Pad\'{e} approximants are similar to those for $l=1-3$
in tables 1 and 2.

In table 6 we display our results along with those from STM and direct
numerical integration (DNI) reported in [35] for the Coulomb plus
logarithmic potential. In table 7 we list our predictions for the cutoff -
Coulomb and compare them with those of Singh et al. [17]. Finally, we report
the k=0 eigenvalues for the NPO and cutoff - Coulomb potentials in table 8
and 9, respectively.

\section{Summary and Remarks}

In this work we have introduced a pseudoperturbative shifted - $l$ expansion
technique (PSLET) to deal with the calculation of the energy eigenvalues and
eigenfunctions of Schr\"odinger equation in one batch. We have shown that it
is an easy task to implement PSLET without having to worry about ranges of
couplings and forms of perturbations in the potential involved. In contrast
to the textbook Rayleigh - Schr\"odinger perturbation theory, an easy
feasibility of computation of the eigenvalues and eigenfuctions has been
demonstrated, and satisfactory accuracies have been obtained. Perhaps it
should be noted that for each entry in tables 1-5 one can construct the
wavefunction from the knowledge of $C_{m,n}$ and $D_{m,n}$. However, such a
study lies beyond the scope of our methodical proposal.

The conclusion reached by Scherrer [2], via MCF method for the NPO potential
Eq.(1), that SLNT [8] leads to dubiously accurate results in the critical
range $a^{1/2}/b\approx 0.1-30$ has been confirmed in the present work.
However, the natural extensions of SLNT or its variants [8,16-21],
represented by PSLET or the modified SLNT [22,23], show that one could get
convincingly reliable results. Moreover, the dubious accuracies of SLNT [8]
should be attributed mainly to the limited capability of SLNT to calculate
the energy series beyond the fourth - order term. Our results for the cutoff
- Coulomb and Coulomb plus logarithmic potentials are also very satisfactory.

A final remark is in point. The attendant method could be applied to systems
at lower dimensions. Here is the recipe. Rewrite the centrifugal term in
Eq.(2) as $\Lambda (\Lambda +1)/2q^{2}$. Shift $\Lambda $ through $\bar{%
\Lambda}=\Lambda -\beta $ and expand in inverse powers of $\bar{\Lambda}$
following exactly the same procedure described in section 2. In this case, $%
\Lambda =l$ in three dimensions ( with $q\geq 0$), $\Lambda =m-1/2$ in two
dimensions where $m$ is the magnetic quantum number ( with $q\geq 0$), and $%
\Lambda =-1$ and/or 0 in one dimension ( with $-\infty <q<\infty $) [23-28].

\newpage

\begin{table}[tbp]
\caption{ Bound - state energies of the NPO potential (1). Where $E_P$
represents PSLET results, Eq.(51), $E_{SS}$ from SSQM [12], $E_E$ from EMM
[14]. The Pad\'{e} approximant $E[3,4]$ and the upper bound of EMM are
obtained, respectively, by replacing the last $j$ digits of $E[3,3]$ and of
the lower bound of EMM with the $j$ digits in parentheses.}
\begin{center}
\vspace{0.5cm} 
\begin{tabular}{|c|c|c|c|c|c|c|}
\hline\hline
$l$ & $a$ & $b$ & $E_P$ & $E[3,3]~\&~(E[3,4])$ & $E_{SS}$ & $E_E$ \\ \hline
1 & 0.1 & 0.1 & 5.1863731 & 5.1863730 (29) & 5.186338 & 5.1863730 (30) \\ 
&  & 0.5 & 5.100883 & 5.100858 (61) & 5.100976 & 5.100842 (65) \\ 
&  & 1 & 5.065556 & 5.065563 (71) & 5.065610 & 5.06428 (609) \\ 
& 0.5 & 0.1 & 5.8935959 & 5.8935951 (27) & 5.893494 & 5.8935952 (52) \\ 
& 1 & 0.1 & 6.7042393 & 6.704239 (46) & 6.704090 & 6.7042389 (89) \\ 
&  & 1 & 5.65138 & 5.65113 (40) & 5.652112 & 5.6503 (21) \\ 
& 10 & 0.1 & 15.8137089 & 15.81370943 (48) & 15.813628 & 15.81370943 (43) \\ 
& 100 & 0.1 & 49.38979430 & 49.389794297 (97) & 49.389615 & 49.38979427 (34)
\\ 
&  & 10 & 14.371 & 14.3622 (64) & 14.363739 & 1.7 (14.609) \\ 
&  & 100 & 5.9934347 & 5.993450 (59) & 5.993565 & ---- (6.389) \\ \hline
2 & 0.1 & 0.1 & 7.24396166 & 7.243961847 (50) & 7.243927 & 7.243961840 (40)
\\ 
&  & 0.5 & 7.118983 & 7.1189816 (10) & 7.119005 & 7.11890 (901) \\ 
&  & 1 & 7.0737228 & 7.0737258 (70) & 7.073713 & 7.0730 (44) \\ 
& 0.5 & 0.1 & 8.17787177 & 8.17787168 (66) & 8.177754 & 8.17787169 (69) \\ 
& 1 & 0.1 & 9.2619150 & 9.26191476 (42) & 9.261812 & 9.26191478 (78) \\ 
&  & 1 & 7.73479 & 7.734821 (33) & 7.734778 & 7.734 (36) \\ 
& 10 & 0.1 & 21.83609251 & 21.83609251 (58) & 21.836043 & 21.83609247 (54)
\\ 
& 100 & 0.1 & 68.802061155 & 68.802061155 (55) & 68.801562 & 68.8020606 (15)
\\ 
&  & 10 & 16.61083 & 16.61081 (54) & 16.611028 & 16.5997 (6540) \\ 
&  & 100 & 7.9960247 & 7.99602475 (91) & 7.996048 & 7.9947 (8.0378) \\ 
\hline\hline
\end{tabular}
\end{center}
\end{table}

\newpage

\begin{table}[tbp]
\caption{ Same as table 1.}
\begin{center}
\vspace{0.5cm} 
\begin{tabular}{|c|c|c|c|c|c|}
\hline\hline
$l$ & $a$ & $b$ & $E_P$ & $E[3,3]~\&~(E[3,4])$ & $E_E$ \\ \hline
3 & 0.1 & 0.1 & 9.2943590 & 9.294359116 (16) & 9.2943591109 (09) \\ 
&  & 0.5 & 9.1318120 & 9.131807 (12) & 9.131799 (838) \\ 
&  & 1 & 9.0789113 & 9.0789116 (13) & 9.0787 (92) \\ 
& 0.5 & 0.1 & 10.4292039 & 10.429204129 (33) & 10.42920412 (12) \\ 
& 1 & 0.1 & 11.7606210 & 11.760620955 (50) & 11.76062096 (96) \\ 
&  & 1 & 9.787665 & 9.787668 (73) & 9.7875 (81) \\ 
& 10 & 0.1 & 27.68830286 & 27.6883029 (31) & 27.6883028 (28) \\ 
& 100 & 0.1 & 88.01806590 & 88.018065906 (07) & 88.0180658 (60) \\ 
&  & 10 & 18.719989 & 18.720001 (21) & 18.7186 (307) \\ 
&  & 100 & 9.99715366 & 9.997153642 (55) & 9.9969 (10.0113) \\ \hline
5 & 200 & 0.1 & 179.48311321 & 179.48311321 (21) & 179.483107 (16) \\ 
& 500 & 0.1 & 286.13076490 & 286.130764905 (05) & 286.13073 (82) \\ 
& 1000 & 0.2 & 401.608033488 & 401.6080334886 (89) & 401.6078 (81) \\ 
& 10000 & 0.4 & 1280.6255249 & 1280.62552494 (95) & 1280.6254 (56) \\ 
& 10000 & 0.5 & 1275.7839677 & 1275.7839677 (77) & 1275.7838 (42) \\ \hline
10 & 200 & 0.1 & 311.86088089 & 311.86088089 (89) & 311.8601371 (16266) \\ 
& 1000 & 0.1 & 713.36153440 & 713.36153440 (40) & 713.36081 (321) \\ 
& 1000 & 0.2 & 699.10562257 & 699.10562257 (57) & 699.10424 (909) \\ 
& 10000 & 0.4 & 2242.79417589 & 2242.79417589 (89) & 2242.7891995 (3867) \\ 
& 10000 & 0.5 & 2228.5184345 & 2228.5184345 (45) & 2228.513255 (30746) \\ 
\hline
20 & 500 & 0.1 & 914.36631099 & 914.36631099 (09) & 914.36540 (851) \\ 
& 1000 & 0.1 & 1312.251674809 & 1312.251674809 (09) & 1312.25006 (333) \\ 
\hline\hline
\end{tabular}
\end{center}
\end{table}

\newpage

\begin{table}[tbp]
\caption{Bound - state energies of the NPO potential (1) for $a=10$. $E_P$, $%
E[3,3]$, and $E[3,4]$ are the same as in table 1. $E_M$ from MCF [2] and $%
E_{SL}$ from SLNT [8].}
\begin{center}
\vspace{0.5cm} 
\begin{tabular}{|c|c|c|c|c|c|}
\hline\hline
$l$ & $b$ & $E_P$ & $E[3,3]~\&~(E[3,4])$ & $E_M$ & $E_{SL}$ \\ \hline\hline
0 & 1 & 7.41837 & 7.417532 (12) & 7.417506 & 7.4056 \\ 
& 10 & 3.885 & 3.864 (75) & 3.879037 & 3.8732 \\ 
& 100 & 3.09826 & 3.097906 (26) & 3.089317 & 3.0984 \\ 
& 1000 & 3.0099813 & 3.0099799 (801) & 3.009981 & 3.0100 \\ \hline
1 & 1 & 11.0714 & 11.07326 (27) & 11.073300 & 11.0714 \\ 
& 10 & 5.94089 & 5.94057 (23) & 5.940860 & 5.9408 \\ 
& 100 & 5.099344 & 5.0993455 (64) & 5.099344 & 5.0994 \\ 
& 1000 & 5.009993354 & 5.009993346 (41) & 5.009993 & 5.0100 \\ \hline
2 & 1 & 14.0861 & 14.08540 (31) & 14.085383 & 14.0900 \\ 
& 10 & 7.9622256 & 7.962228 (57) & 7.962230 & 7.9622 \\ 
& 100 & 7.09960269 & 7.099602614 (30) & 7.099603 & 7.0996 \\ 
& 1000 & 7.009996003 & 7.00999600266 (38) & 7.009996 & 7.0100 \\ \hline
3 & 1 & 16.719284 & 16.71910 (33) & 16.719332 & 16.7200 \\ 
& 10 & 9.9724544 & 9.9724556 (62) & 9.972455 & 9.9724 \\ 
& 100 & 9.09971542 & 9.09971542 (42) & 9.099715 & 9.0998 \\ 
& 1000 & 9.00999714406 & 9.00999714399 (92) & 9.009997 & 9.0100 \\ \hline
4 & 1 & 19.137789 & 19.137816 (24) & 19.137821 & 19.1376 \\ 
& 10 & 11.97836435 & 11.97836459 (62) & 11.978365 & 11.9784 \\ 
& 100 & 11.099778408 & 11.099778407 (07) & 11.099778 & 11.0998 \\ 
& 1000 & 11.00999777842 & 11.009997778412 (64) & 11.009998 & 11.0100 \\ 
\hline\hline
\end{tabular}
\end{center}
\end{table}

\newpage

\begin{table}[tbp]
\caption{ Same as table 3 for $a=100$.}
\begin{center}
\vspace{0.5cm} 
\begin{tabular}{|c|c|c|c|c|c|}
\hline\hline
$l$ & $b$ & $E_P$ & $E[3,3]~\&~(E[3,4])$ & $E_M$ & $E_{SL}$ \\ \hline\hline
0 & 1 & 26.70597 & 26.705966 (75) & 26.705966 & 26.706 \\ 
& 10 & 10.6 & 11.55 (76) & 11.572197 & 11.6112 \\ 
& 100 & 3.9825 & 3.9790 (17) & 3.983098 & 3.9844 \\ 
& 1000 & 3.099813 & 3.0997996 (8014) & 3.099811 & 3.0998 \\ \hline
1 & 1 & 42.23757 & 42.2375612 (88) & 42.237560 & 42.236 \\ 
& 10 & 14.370 & 14.3622 (64) & 14.368811 & 14.3638 \\ 
& 100 & 5.9934346 & 5.9934500 (91) & 5.993439 & 5.9936 \\ 
& 1000 & 5.09993354 & 5.09993346 (41) & 5.099933 & 5.1000 \\ \hline
2 & 1 & 55.97780 & 55.9778047 (70) & 55.977804 & 55.976 \\ 
& 10 & 16.61084 & 16.61081 (54) & 16.610869 & 16.6110 \\ 
& 100 & 7.99602470 & 7.99602475 (91) & 7.996025 & 7.9960 \\ 
& 1000 & 7.099960046 & 7.0999600264 (37) & 7.099960 & 7.1000 \\ \hline
3 & 1 & 67.9608094 & 67.96080 (81) & 64.960806 & 67.9600 \\ 
& 10 & 18.719989 & 18.720001 (21) & 18.719999 & 18.7202 \\ 
& 100 & 9.997153662 & 9.997153641 (55) & 9.997154 & 9.9972 \\ 
& 1000 & 9.0999714406 & 9.099971440 (39) & 9.099971 & 9.1000 \\ \hline
4 & 1 & 78.2383822 & 78.23838049 (45) & 78.238380 & 78.2380 \\ 
& 10 & 20.7814134 & 20.7814161 (67) & 20.781416 & 20.7820 \\ 
& 100 & 11.9977838 & 11.9977838311 (34) & 11.997784 & 11.9978 \\ 
& 1000 & 11.099977784 & 11.099977784 (84) & 11.099978 & 11.10000 \\ 
\hline\hline
\end{tabular}
\end{center}
\end{table}

\newpage

\begin{table}[tbp]
\caption{ Same as table 3 for $a=1000$.}
\begin{center}
\vspace{0.5cm} 
\begin{tabular}{|c|c|c|c|c|c|}
\hline\hline
$l$ & $b$ & $E_P$ & $E[3,3]~\&~(E[3,4])$ & $E_M$ & $E_{SL}$ \\ \hline\hline
0 & 1 & 91.2566 & 91.25661112 (32) & 91.256611 & 91.2560 \\ 
& 10 & 64.833 & 64.8244 (51) & 64.825083 & 64.7400 \\ 
& 100 & 12.8162 & 12.784 (80) & 12.823345 & 12.8366 \\ 
& 1000 & 3.99813 & 3.99800 (01) & 3.998107 & 3.9984 \\ \hline
1 & 1 & 149.6563194 & 149.65631949 (81) & 149.656319 & 149.6560 \\ 
& 10 & 89.126 & 89.12349 (37) & 89.123452 & 89.0400 \\ 
& 100 & 14.933728 & 14.93398 (406) & 14.933774 & 14.9350 \\ 
& 1000 & 5.9993353 & 5.99933458 (11) & 5.999335 & 5.9994 \\ \hline
2 & 1 & 206.1068055 & 206.1068055 (59) & 206.106805 & 206.10000 \\ 
& 10 & 100.60 & 100.7043 (64) & 100.703996 & 100.7260 \\ 
& 100 & 16.9601067 & 16.9601077 (91) & 16.960106 & 16.9604 \\ 
& 1000 & 7.99960031 & 7.999600250 (23) & 7.999600 & 7.99960 \\ \hline
3 & 1 & 260.6091863 & 260.6091863 (71) & 260.609186 & 260.6200 \\ 
& 10 & 105.5097 & 105.5054 (70) & 105.507579 & 105.4960 \\ 
& 100 & 18.9714849 & 18.97148472 (84) & 18.971485 & 18.9716 \\ 
& 1000 & 9.99971440 & 9.999714394 (86) & 9.999714 & 9.9998 \\ \hline
4 & 1 & 313.16466655 & 313.164666 (72) & 313.164667 & 313.1600 \\ 
& 10 & 108.52780 & 108.527802 (18) & 108.527834 & 108.5280 \\ 
& 100 & 20.9778139 & 20.977813876 (97) & 20.977814 & 20.9780 \\ 
& 1000 & 11.999777840 & 11.999777839 (44) & 11.999778 & 11.9998 \\ 
\hline\hline
\end{tabular}
\end{center}
\end{table}

\newpage

\begin{table}[tbp]
\caption{$k=0$ eigenvalues for $V(q)=-1/2q+(\protect\mu/2) ln(q^2+q)$.}
\begin{center}
\vspace{0.5cm} 
\begin{tabular}{|c|c|c|c|c|}
\hline\hline
$\mu $ & $E_P$ & E[4,4] & STM & DNI \\ \hline
0.0001 & 0.2497779 & 0.2497779 & 0.24975 & 0.24978 \\ 
0.0005 & 0.248890 & 0.248890 & 0.24875 & 0.24889 \\ 
0.001 & 0.247782 & 0.247782 & 0.24752 & 0.24778 \\ 
0.005 & 0.238976 & 0.238973 & 0.23765 & 0.23897 \\ 
0.01 & 0.228105 & 0.228098 & 0.22545 & 0.28810 \\ 
0.05 & 0.145676 & 0.145681 & 0.13227 & 0.14568 \\ 
0.1 & 0.051363 & 0.051499 & 0.02456 & 0.05153 \\ 
0.5 & 0.521641 & 0.520529 & 0.65413 & 0.52033 \\ \hline
\end{tabular}
\end{center}
\end{table}

\begin{table}[tbp]
\caption{CUTT OFF POTENTIAL.}
\begin{center}
\vspace{0.5cm} 
\begin{tabular}{|c|c|c|c|c|c|}
\hline\hline
$b$ &  & $1s$ & $2p$ & $3d$ & $4f$ \\ \hline
0.3 & $E_P$ & 0.2938592647 & 0.1061734109 & 0.0516188327 & 0.0299960986 \\ 
& $E[4,4]$ & 0.2935835059 & 0.1061736042 & 0.0516188327 & 0.0299960986 \\ 
& $E_{ex}$ & 0.29354528 & 0.10617351 & 0.05161883 & 0.02999610 \\ 
0.5 &  & 0.2446947653 & 0.0976555780 & 0.0515016542 & 0.0292431052 \\ 
&  & 0.2445430901 & 0.0976556895 & 0.0515016549 & 0.0292431052 \\ 
&  & 0.24453144 & 0.09765562 & 0.04943696 & 0.02924311 \\ 
5 &  & 0.0706714094 & 0.0434584151 & 0.0287056893 & 0.0200000000 \\ 
&  & 0.0706701591 & 0.0434584053 & 0.0287056891 & 0.0200000000 \\ 
&  & 0.07067028 & 0.04345840 & 0.02870569 & 0.02000000 \\ 
100 &  & 0.0067420934 & 0.0056337380 & 0.0048124934 & 0.0041687844 \\ 
&  & 0.0067420738 & 0.0056337379 & 0.0048124934 & 0.0041687844 \\ 
&  & 0.00674208 & 0.00563374 & 0.00481249 & 0.00416878 \\ \hline\hline
\end{tabular}
\end{center}
\end{table}

\begin{table}[tbp]
\caption{NPO for $k=1$ and $a=10$.}
\begin{center}
\vspace{0.5cm} 
\begin{tabular}{|cccccc|}
\hline\hline
$b$ & $l$ & $E_{PSLET}$ & $E_{44}$ & $E_{1/N}$ & $E_{CFM}$ \\ \hline\hline
$1$ & $0$ & $13.3557146$ & $13.388417$ & $13.496$ & $13.388323$ \\ 
$10$ &  & $7.9029106$ & $7.9654956$ & $7.9180$ & $7.903154$ \\ 
$100$ &  & $7.0984744$ & $7.0979219$ & $7.0990$ & $7.098449$ \\ 
$1000$ &  & $7.0099838$ & $7.0099799$ & $7.0100$ & $7.009982$ \\ \hline
$1$ & $1$ & $16.0066200$ & $16.0156839$ & $15.963$ & $16.016128$ \\ 
$10$ &  & $9.9447832$ & $9.9452031$ & $9.9504$ & $9.944898$ \\ 
$100$ &  & $9.0993550$ & $9.0993626$ & $9.0994$ & $9.099352$ \\ 
$1000$ &  & $9.0099934$ & $9.0099934$ & $9.0100$ & $9.009993$ \\ \hline
$1$ & $2$ & $18.5442860$ & $18.5432358$ & $18.498$ & $18.543473$ \\ 
$10$ &  & $11.9633289$ & $11.9633409$ & $11.965$ & $11.963343$ \\ 
$100$ &  & $11.0996043$ & $11.0996041$ & $11.010$ & $11.099604$ \\ 
$1000$ &  & $11.0099960$ & $11.0099960$ & $11.010$ & $11.009996$ \\ \hline
\end{tabular}
\end{center}
\end{table}

\newpage

\end{document}